# Effect of Fe substitution on the magnetic, transport, thermal and magnetocaloric properties in $Ni_{50}Mn_{38-x}Fe_xSb_{12}$ Heusler alloys

Roshnee Sahoo[1], Ajaya K. Nayak[1], K. G. Suresh[1,*] and A. K. Nigam[2]

[1]Magnetic Materials Laboratory, Department of Physics, Indian Institute of Technology Bombay, Mumbai- 400076, India

[2]Tata Institute of Fundamental Research, Homi Bhabha Road, Mumbai- 400005, India

**Abstract**

The structural, magnetic, transport, thermal and magnetothermal properties of quaternary Heusler alloys $Ni_{50}Mn_{38-x}Fe_xSb_{12}$ have been studied. Powder x-ray diffraction and temperature dependence of magnetization studies reveal that with addition of Fe in Mn site, the martensitic transition shifts to low temperatures. It is also found that the martensitic transition becomes broader with increase in Fe concentration. The metamagnetic transition in *M(H)* isotherms becomes very prominent in $x=2$ and vanishes for $x=3$ and 4. A maximum positive magnetic entropy change of 14.2 J/kg K is observed for $x=2$ at 288 K for 50 kOe. Electrical resistivity data show an abrupt decrease across the martensitic transition in all the alloys, except $x=6$, which does not have the martensitic transition. A maximum negative magnetoresistance of 21% has been obtained for $x=2$ at 50 kOe. The same alloy also shows an exchange bias field of 288 Oe.

**Keywords:** Heusler alloys, martensitic transition, magnetocaloric effect, exchange bias

*Corresponding author (email: suresh@phy.iitb.ac.in, FAX: +91-22-25723480)

# 1. Introduction

The full Heusler alloys, particularly NiMn based alloys, have attracted much attention owing to their applications in various fields owing to their interesting multi-functional properties like shape memory effect,[1,2] magnetocaloric effect (MCE),[3,4] magnetoresistance (MR)[5,6] and exchange bias (EB) behaviour[7,8] etc. These properties mainly arise due to the martensitic transition, which in many cases is coupled with the magnetic transition, giving rise to a first order magneto-structural transition from an ordered austenite (cubic) phase to a less ordered martensite (tetragonal or orthorhombic) phase on cooling. The magnetic properties of NiMn based alloys can be mainly attributed to the Mn magnetic moments since Ni atoms have almost negligible moment.[9] The Mn-Mn exchange interaction, which depends on the Mn-Mn bond length, plays an important role in determining the magnetism in these alloys. In their stoichiometric form, $Ni_2MnZ$ (Z=In, Sn and Sb) alloys do not exhibit any martensitic transition. When Z is partially replaced by Mn, martensitic transition is observed for some critical concentrations. For example, in $Ni_{50}Mn_{25+x}Sb_{25-x}$, for certain values of $x$ ($7 \leq x \leq 10$), the system exhibits martensitic transition.[10] This transition, which can be induced by changing the chemical composition, can also be tuned with the application of hydrostatic pressure as well as applied magnetic field.[11] The magnetism in these alloys mainly arises due to the RKKY exchange interaction.[12]

Recently Ni-Mn-Sb alloys have been receiving a lot of interest due to their anomalous magnetic properties associated with the magneto-structural transition.[13,14] Inverse magnetocaloric effect (IMCE) i.e., the positive magnetic entropy change ($\Delta S_M^{max}$) of 7.6 J/kg K, 12.1 J/kg K, 21.9 J/kg K and 6.1 J/kg K in $Ni_{50-x}Mn_{38+x}Sb_{12}$ alloys for x = -1, 0, 1 and 2 respectively have been reported by Feng et. al.[15] Furthermore, the exchange bias field up to 248 Oe has been reported in Ni-Mn-Sb alloys after cooling in a field of 50 kOe.[7] The exchange bias behaviour has been attributed to the coexistence of ferromagnetic (FM) and antiferromagnetic (AFM) components in the system.[16] It was recently shown that the martensitic transition temperature as well as the magnetization of the Ni-Mn-Sb system can be significantly changed by substitution of Co at the Ni site, which gives rise to a large positive MCE of 34 J/kg K in the martensitic transition region and also a large exchange bias of 480 Oe at low temperatures.[8,17] It is also reported that in $Ni_2(Fe,Mn)Ga$, Curie temperature and magnetization can be enhanced by the substitution of Fe.[18] Using Mossbauer studies in the same system, it was found that substituting Fe at the Mn site results in higher saturation magnetization and improved exchange interaction in the Mn sublattice.[19] In view of these, we decided to study the effect of Fe substitution for Mn in $Ni_{50}Mn_{38-x}Fe_xSb_{12}$ series. We also find that such a study has not been carried out in any NiMnSb based system. In this paper, we report the structural, magnetic, magnetotransport, and magnetothermal properties of $Ni_{50}Mn_{38-x}Fe_xSb_{12}$ ($1 \leq x \leq 6$) alloys.

## 2. Experimental details

Alloy ingots of $Ni_{50}Mn_{38-x}Fe_xSb_{12}$ were prepared by arc melting the stoichiometric amounts of Ni, Fe, Mn and Sb of atleast 99.99% purity in argon atmosphere. Approximately 2% extra Mn was added to compensate the weight loss due to the high vapour pressure of Mn. The ingots were remelted three times and the final weight loss was found to be negligible. The as-cast ingots were annealed at 850° C for 24 hours in an evacuated quartz tube for homogenization and then slowly cooled. The structural characterization was done by powder x-ray diffraction (XRD) using Cu-Kα radiation.

The magnetization measurements were carried out using a vibrating sample magnetometer attached to a Physical Property Measurement System (Quantum Design, PPMS-6500) and the electrical resistivity ($\rho$) measurements were performed by four probe method using PPMS. The heat capacity ($C_P$) measurement was also done using the PPMS.

## 3. Results and Discussion

Powder x-ray diffraction patterns of $Ni_{50}Mn_{38-x}Fe_xSb_{12}$ collected at room temperature are shown in figure 1, which show that all the compounds except $x=1$ are in the cubic (austenite) phase with the $L2_1$ Heusler structure at room temperature. As can be seen from the inset of figure 1(a), two satellite peaks exist in the (220) peak for $x=1$. This indicates the presence of some martensitic feature in that alloy. It may be noted that $Ni_{50}Mn_{38}Sb_{12}$ possesses the martensite phase at room temperature.[8,10] With increase in the $x$ value, the intensity of the satellite peaks decreases considerably and above $x=2$ the structure is purely austenite at room temperature. The typical Rietveld refinement is shown in figure 1(e) for $x=5$, which is completely in the single austenite phase. Therefore, it can be seen that by substituting Fe in Mn site in $Ni_{50}Mn_{38}Sb_{12}$, the room temperature phase changes to austenite, thereby indicating the stabilization of the austenite phase with Fe substitution.

**Table 1**. Transition temperatures, (*e/a*) ratio and magnetic entropy change (*ΔS$_M$*) as a function of *x* in Ni$_{50}$Mn$_{38-x}$Fe$_x$Sb$_{12}$

| x | M$_S$ (K) | M$_F$ (K) | A$_S$(K) | A$_F$ (K) | $T_M(K) = \frac{(M_S + M_F)}{2}$ | (e/a) | (ΔS$_M$)$_{max}$ (J/ kg K) |
|---|---|---|---|---|---|---|---|
| 0 | 328 | 315 | 330 | 340 | 321.5 | 8.26 | 7 |
| 1 | 279 | 273 | 288 | 295 | 276 | 8.27 | 6.3 |
| 2 | 278 | 273 | 290 | 294.8 | 275.5 | 8.28 | 14.2 |
| 3 | 199 | 189 | 207.5 | 217 | 194 | 8.29 | 4.4 |
| 4 | 214 | 159 | 179 | 230 | 186.5 | 8.30 | -- |
| 5 | 167.5 | 32 | 36 | 190.5 | 100 | 8.31 | -- |
| 6 | 124 | 47 | 23.5 | 136 | 85.5 | 8.32 | |

The temperature dependence of magnetization, *M(T)*, of Ni$_{50}$Mn$_{38-x}$Fe$_x$Sb$_{12}$ compounds is shown in figure 2 (a) to (f). The measurement has been performed in two different modes. In the ZFC mode the sample was cooled from 330 to 5 K without any field and the data were taken by heating from 5 to 330 K after applying a field of 1 kOe. In the FCC mode, the data were taken by cooling the sample from 330 to 5 K in the same field. In figure 2(a), following the FCC curve, it can be seen that at high temperatures, there is a transition from paramagnetic to ferromagnetic phase at the Curie temperature ($T_C^A$) of the austenite phase. On further decreasing the temperature, the alloy undergoes the martensitic transition, which starts at *M$_S$* (where *M(T)* curve reaches the maximum) and ends at *M$_F$* (where *M(T)* curve reaches the minimum). Temperature corresponding to the martensitic transition is defined as the mean of the martensitic start and finish temperatures, *i.e.*, $T^M = (M_S + M_F)/2$. Below *M$_F$*, with decrease in temperature there is another transition at $T_C^M$, which signifies the Curie temperature of the martensite phase. In a similar manner, the reverse transformation occurring on heating defines the austenite start ($A_S$) and austenite finish ($A_F$) temperatures. Different characteristic temperatures defining the martensitic transition are listed in Table 1. From the table it can be noticed that the martensitic transition temperatures monotonically decrease with increase in Fe concentration. There have been reports that the $T^M$ increases with increase in the average valence electrons per atom ($e/a$).[20,21] However, in the present case, it seems that such a relationship does not hold good as the $T^M$ and the ($e/a$) ratio show opposite trends. Therefore, it is clear that at least in certain systems, the factor that drives the martensitic transition cannot be simply described in terms of ($e/a$). The large magneto-structural coupling brought about by Fe plays a role in driving the martensitic transition. It is also evident that the martensitic transition becomes broader with increase in $x$ and it almost disappears above $x=6$. This may be due to the enhancement of ferromagnetic coupling with Fe substitution, which suppresses the AFM component in the martensite phase. This results in the stabilization of the austenite phase and leads to the shifting of the martensitic transition towards lower temperatures.

From the Fig. 2, it may also be noted that both $T_C^A$ and $T_C^M$ depend on the $x$ value. Another interesting point to be noted is that the ferro to paramagnetic transition of the martensite phase (at $T_C^M$) is absent for $x>2$. In all the alloys, the bifurcation between the ZFC and the FCC curves is present at low temperatures. It is well known from the results on NiMnSn alloys that in non-stoichiometric alloys, the extra Mn occupies the Sn site and that the coupling between the regular Mn and the Mn at the Sn site is antiferromagnetic.[22] By lowering the temperature, this coupling strengthens in the martensite phase.[23] A similar scenario may be expected in NiMnSb as well. As mentioned earlier, substitution of Fe for Mn would suppress the AFM interaction, which in turn destabilizes the martensite phase. The splitting between the ZFC and FCC curves was also observed in NiMnSb system.[10] It may arise due to the presence of both AFM and FM interactions in the system. With increase in field, the splitting disappears (as seen in inset of fig. 2(b)) as a field of 50 kOe is sufficient to overcome the AFM component. A similar behaviour in *M (T)* has also been observed in $Ni_{50}(Mn_{1-x}Fe_x)_{36}Sn_{14}$.[24]

In figure 3, field dependence of magnetization, *M(H)*, is shown for x = 1, 2, 3 and 4 for $Ni_{50}Mn_{38-x}Fe_xSb_{12}$ alloys in the martensitic transition region. For *x* = 1, *M(H)* measured in 290 K $\leq T \leq$ 300 K shows the ferromagnetic behaviour. However, with increase in temperature, the magnetization is found to increase. Field induced transition *i.e.,* metamagnetic transition can be seen from 292 K to 298 K at about 33 kOe. Such a behaviour vanishes at higher temperatures. A similar behaviour is also observed in figure 3(b) for *x*=2, where the isotherms have been taken in the temperature range of 285 K to 290 K. Here the metamagnetic transition is observed at 287, 288 and 289 K. It may be noted that the metamagnetic transition in *x*=2 is sharper than that of *x*=1. This is reflective of the fact that the martensitic transition is sharper in the former, as is also evident from the *M(T)* plot in figure 2. For *x* = 3, though the magnetization increases with increasing temperature, there is no sign of a metamagnetic transition upto 50 kOe. For x=4 there is only a marginal increase in the magnetization with temperature. Comparing all the compounds, it is seen that the magnetization of the austenite phase shows a monotonic increase with *x*. This implies that the Fe substitution enhances the ferromagnetic coupling in this system. Consequently the martensite phase loses its stability and the system tends to behave like a normal ferromagnet (austenite state). This is in agreement with the fact that the martensitic transition gradually vanishes with Fe addition. The enhanced FM behaviour is also seen in the saturation observed in the *M(H)* plot at 50 kOe. For *x*=3 and 4, the saturation tendency is better than that in *x*=1 and 2.

The insets of figure 3(a) and (b) show the hysteresis loss (HL) observed at different temperatures in the martensitic transition region. For x=1 and x=2, the average hysteresis loss around the martensitic transition region is found to be to be 4.6 and 8 J/kg respectively. These values are found to be significantly lower compared to other Heusler alloys.[13,25]

As the temperature variation of magnetization is nearly continuous, Maxwell's relation can be used to get an approximate estimate of the magnetic entropy change in the vicinity of the martensitic region, though strictly its use for first order systems is not correct.[26,27] Recently in $Ni_{50-x}Co_xMn_{32-y}Fe_yGa_{18}$ system, a large MCE of 31 J/Kg K has been reported near the first order transition region using the Maxwell relation.[28] Figure 4 shows the inverse magnetocaloric effect of $Ni_{50}Mn_{38-x}Fe_xSb_{12}$ evaluated in the region of the martensitic transition. Maximum positive magnetic entropy change values are obtained as 6.3 J/kg K for *x*=1, 14.2 J/kg K for *x*=2 and 4.4 J/kg K for *x*=3 in 50 kOe. As the use of Maxwell's relation for the first order systems is still questionable, we have also evaluated the magnetic entropy change in the decreasing field mode, which is shown by the dotted curve in figure 4(a) and (b). It can be seen that the value of $\Delta S_M$ is found to be 6.3 J/kg K for x=1 and 13.6 J/kg K for x=2 in this case. Therefore, the difference between the increasing field and decreasing field modes is quite nominal. As expected, the peak position shifts to low temperatures in the field decreasing mode. It is reported that the maximum *ΔS_M* value calculated using the Maxwell's relation for $Ni_{50}Mn_{38}Sb_{12}$ is ~7 J/kg K in 50 kOe.[17] Therefore, it is clear that the MCE variation as a function of Fe concentration goes through a peak, with the maximum occurring at $x=2$. The large increase in $x=2$ is due to the large $(\partial M/\partial T)$, which results from the enhanced magneto-structural coupling brought about by Fe. Furthermore, it may be noted that the temperature corresponding to the maximum entropy change decreases from 295.5 K to 208.5 K as $x$ is varied from 1 to 3. The decrease in the MCE for $x>2$ can be attributed to the fact that the martensite phase does not have a paramagnetic phase. Therefore, the $(\partial M/\partial T)$ is smaller in $x=3$ than in $x=2$, resulting in a decrease in MCE.

The refrigerant capacity ($RC$) value is calculated by integrating $\Delta S_M(T)$ curve over the full width at half maximum. The effective $RC$ value is estimated from the calculated $RC$ value by subtracting the hysteresis loss. For $x=1$ and 2, the effective $RC$ of 27.8 J/Kg and 27.2 J/Kg have been obtained respectively for a field of 50 kOe. In the case of former, the hysteresis loss is much smaller than that of the latter and hence the effective $RC$ value for both the alloys is comparable.

Figure 5 shows the variation of heat capacity with temperature for $x=2$, 4 and 5 in $Ni_{50}Mn_{38-x}Fe_xSb_{12}$ alloys in zero and 50 kOe. It is known fact that the heat capacity in the solid arises due to the lattice, electronic and magnetic contributions. At low temperatures, the lattice and electronic parts show $T^3$ and $T$ dependencies respectively. However, the heat capacity saturates to a constant value at high temperatures. As shown in the figure, a sharp peak is observed at 286.5 K for $x=2$ near the martensite transition, which is due to the first order nature of the transition. On application of 50 kOe field the temperature corresponding to this peak shifts to a lower value as is shown in the inset of the figure 5(a), which implies that the martensite phase gets gradually destabilized with the field. It is also observed from figure 5 (b) and (c) that the peak height gradually decreases and that the peak broadens with Fe addition. This observation also reflects the change in the order of transition in the system. It may be noted that this feature resembles the thermomagnetic curves shown in figure 2, where the sharpness of the martensitic transition is found to decrease with Fe. A small peak present near 270 K for 0 and 50 kOe for $x=2$, as shown in the inset of figure 5(a), corresponds to the paramagnetic to ferromagnetic transition at $T_C^M$. However this feature is absent for $x=4$, which is consistent with the *M-T* data. In this compound, above the martensite transition, another peak is observed near $T_C^A$ as shown in the inset of figure 5(b). Though such a peak is expected for $x=2$ as well, due to the limited temperature range of the measurement, it is not visible. These weak peaks must be associated with the second order magnetic transition, as also observed in $Gd_5Sb_{0.5}Ge_{3.5}$ single crystals.[29] Therefore, the magnetization and heat capacity data are in agreement with each other in this series.

In order to shed more light on the magnetic state of these alloys near the martensitic transition, we have also performed electrical resistivity measurements. Figure 6 shows the temperature dependence of electrical resistivity in the heating and the cooling modes in zero field. It is observed that at low temperatures the resistivity remains nearly a constant for all the alloys and with increase in temperature there is a sudden decrease in the resistivity at the martensitic transition. This change in the resistivity becomes weaker and the temperature at which the decreases occurs shifts to lower values with increase in Fe concentration. The temperature coefficient of resistivity in the martensite region is very small, which is due to high degree of atomic disorder in the system. The hysteresis observed in these plots between the heating and the cooling modes again confirms the first order nature of the transition.

The temperature variation of magnetoresistance (*MR*) is shown in figure 7 for $x=1$, 2 and 3 in $Ni_{50}Mn_{38-x}Fe_xSb_{12}$ alloys. The *MR* was calculated by using the relation $MR = [\{\rho(H, T) - \rho(0, T)\}/ \rho(0, T)] \times 100$. For the field change of 50 kOe, the maximum *MR* of 10%, 21% and 4% have been obtained for $x=1$, 2 and 3 at 293 K, 289 K and 210 K respectively in the martensitic transition region. The highest *MR* of 21 % observed in $x=2$ is attributed to the large difference in the magnetization between the austenite and the martensite phases. For $x>2$, the *MR* values are considerably less due to destabilization of the martensite phase. It is of interest to note that among the alloys studied here, both the magnetic entropy change and in the magnetoresistance show the highest values at $x=2$.

In order to investigate FM and AFM interaction in the system, exchange bias (*EB*) properties have been studied. For $x=2$ and 4, *M-H* data for the field variation of -20 kOe to +20 kOe have been taken at 5 K, after zero field cooling and field cooling in 50 kOe. Figure 8 shows the ZFC data for -5 to +5 kOe range, which shows a double shifted loop. On the other hand, the field cooled curve shows a shift towards negative fields. The FC *M-H* loop taken at different temperatures is shown in figure 9. The exchange bias *i.e.*, the shift in the hysteresis loop is observed at 5 K. The double shifted ZFC loop and the shifted FC loop clearly indicate the presence of exchange bias arising from the exchange anisotropy associated with the FM/AFM interface.[7] Presence of AFM in the system has already been demonstrated by the thermomagnetic irreversibility in the *M(T)* curve.

The variations of exchange bias field ($H_{EB}$) and the coercivity field ($H_C$) for $x=2$ and 4 with temperature are plotted in figure 10. $H_{EB}$ and $H_C$ are calculated as $H_{EB} = -(H_+ + H_-)/2$ and $H_C = |H_+ - H_-|/2$, where $H_+$ and $H_-$ are the positive and negative fields at which the magnetization becomes zero. The maximum $H_{EB}$ is observed to be 288 Oe and 192 Oe for $x=2$ and 4 respectively. It is observed that with increasing temperature $H_{EB}$ values decrease and become negligible above 60 K for $x=2$ and above 40 K for $x=4$, which are identified as the blocking temperature ($T_B$). The reduction in $H_{EB}$ may be due to the suppression of AFM component at higher temperatures, which weakens the coupling between the FM and AFM phases. The $H_C$ shows an initial increase, reaches a maximum and then decreases with increase in temperature. It may be noted that with increasing Fe concentration from $x=2$ to 4, the exchange bias field decreases. This signifies the weakening of AFM/FM coupling with Fe substitution. The reduction of *EB* can be attributed to the growth of FM clusters.[30] The maximum $H_C$ value also decreases from 578 Oe to 481 Oe as x increases from 2 to 4. It can also be observed that for $x=4$, the saturation magnetization is larger than that of $x=2$. Therefore, the enhancement of FM contribution with Fe substitution for Mn explains the reduction of *EB* as well. Another interesting point that can be noticed here is that a small increase in Fe concentration results in a considerable variation in $H_{EB}$ and $H_C$ in this series.

In summary, substitution of a small amount ($x=2$) of Fe is found to be optimal in enhancing the magnetic and other related properties of these alloys. The enhancement of FM interaction brought about by Fe appears to be maximum at $x=2$, which results in the sharpest martensitic transition resulting in large MCE and MR. For $x<2$, the properties are nearly similar to that of the parent alloy. For $x>2$, the FM interaction has become sufficiently strong so that the martensitic transition becomes broad. However, even at *x*=6, there seems to exist some AFM component at low temperatures, as revealed by the *M-T* data.

## 4. Conclusion

Substitution of Fe for Mn in $Ni_{50}Mn_{38-x}Fe_xSb_{12}$ is found to cause considerable changes in the magnetic, transport, thermal, magnetothermal and exchange bias properties. The martensite phase gets destabilized with Fe and the variation of the martensitic transition temperature is not in accordance with the change in the (*e/a*) ratio. The strength of AFM component in the martensite phase is found to be reduced with the substitution of Fe. For *x*>2, the martensite phase seems to have no paramagnetic phase even at the highest temperature. The magnetocaloric effect and the magnetoresistance values are found to be highest in *x*=2, which also shows a large exchange bias. The enhanced magneto-structural coupling and the reduction in the AFM component as a result of Fe substitution seem to explain the observed variations. On the basis of various results obtained, it is clear that the present series is a promising system for multifunctional applications.

## Acknowledgement

The authors thank D. Buddhikot for his help in the transport measurements.

**Figure Captions:-**

FIG. 1. X-ray diffraction patterns of $Ni_{50}Mn_{38-x}Fe_xSb_{12}$ alloys for $x$ = 1, 2, 3, 4 and 5 at room temperature. The Rietveld refinement is shown for $x$ = 5.

FIG. 2. (a-f) ZFC and FCC magnetization curves as a function of temperature at 1 kOe in $Ni_{50}Mn_{38-x}Fe_xSb_{12}$. The inset in (b) shows the variation in $x$=2 for a field of 50 kOe.

FIG. 3. Isothermal magnetization curves of $Ni_{50}Mn_{38-x}Fe_xSb_{12}$ for $x$ = 1, 2 and 3 at various temperatures in the martensitic transition region. The insets in (a) and (b) show the temperature variation of the hysteresis loss.

FIG. 4. Temperature variation of isothermal magnetic entropy change ($\Delta S_M$) in $Ni_{50}Mn_{38-x}Fe_xSb_{12}$ calculated for the increasing field mode. For $x$=1 and 2, magnetic entropy change has been calculated using the decreasing field data also, as shown by the dashed line.

FIG. 5. Temperature dependence of heat capacity in zero and 50 kOe fields for $x$ =2, 4 and 5 in $Ni_{50}Mn_{38-x}Fe_xSb_{12}$ alloys. The inset shows the shift of the peak with field.

FIG. 6. Electrical resistivity as a function of temperature in cooling and heating modes in $Ni_{50}Mn_{38-x}Fe_xSb_{12}$ for $x$=1, 2 to 6 in zero field.

FIG. 7. Magnetoresistance *vs.* temperature for $x$=1, 2 and 3 in $Ni_{50}Mn_{38-x}Fe_xSb_{12}$ alloys.

FIG. 8. ZFC and FC (at 50 kOe) hysteresis loops of $Ni_{50}Mn_{36}Fe_2Sb_{12}$ at 5K

FIG. 9. *M vs. H* loops for $Ni_{50}Mn_{36}Fe_2Sb_{12}$ at different temperatures after field cooling in 50 kOe.

FIG. 10. $H_{EB}$ and $H_C$ field values at different temperatures in $Ni_{50}Mn_{38-x}Fe_xSb_{12}$ with x=2 and 4.

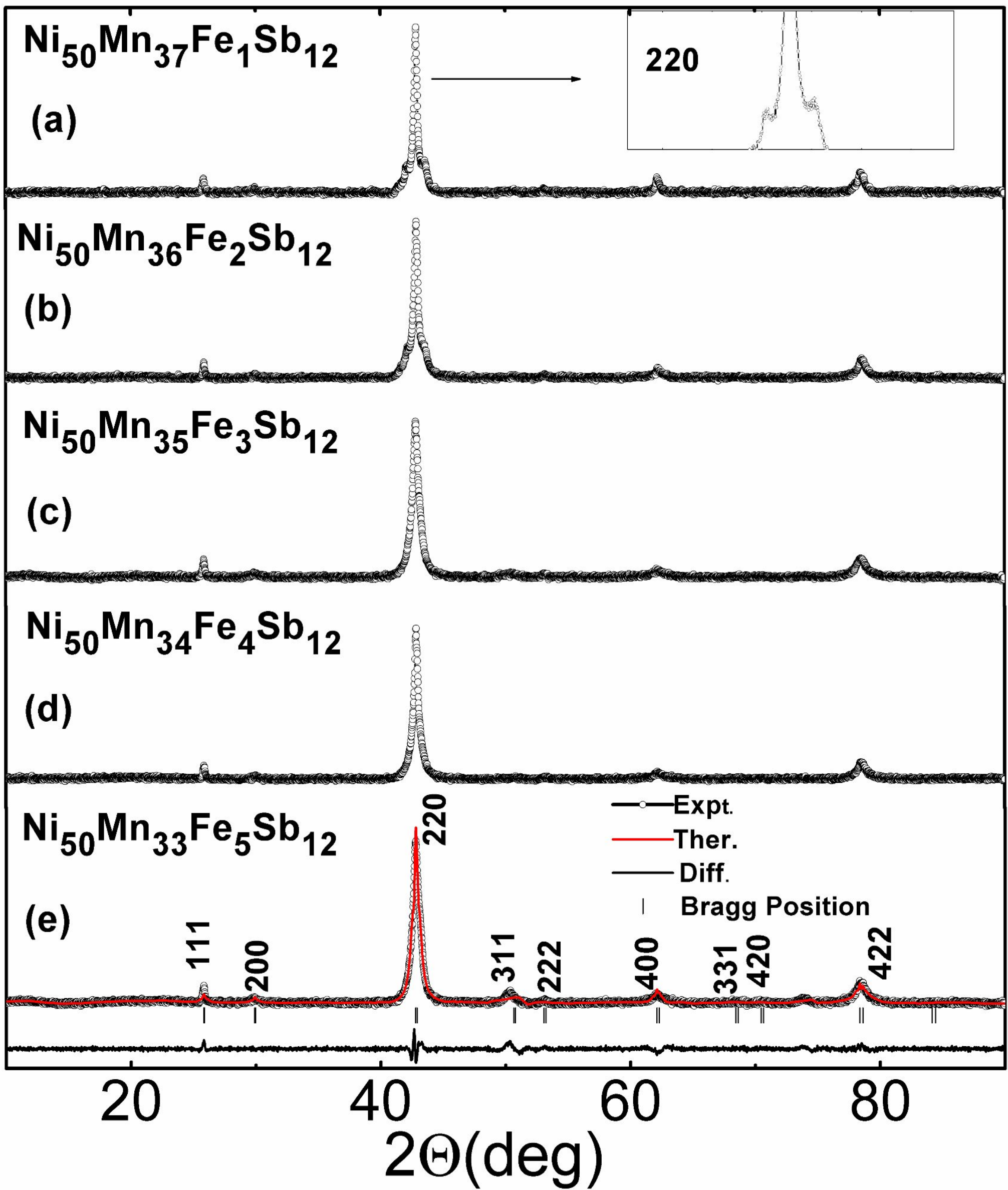

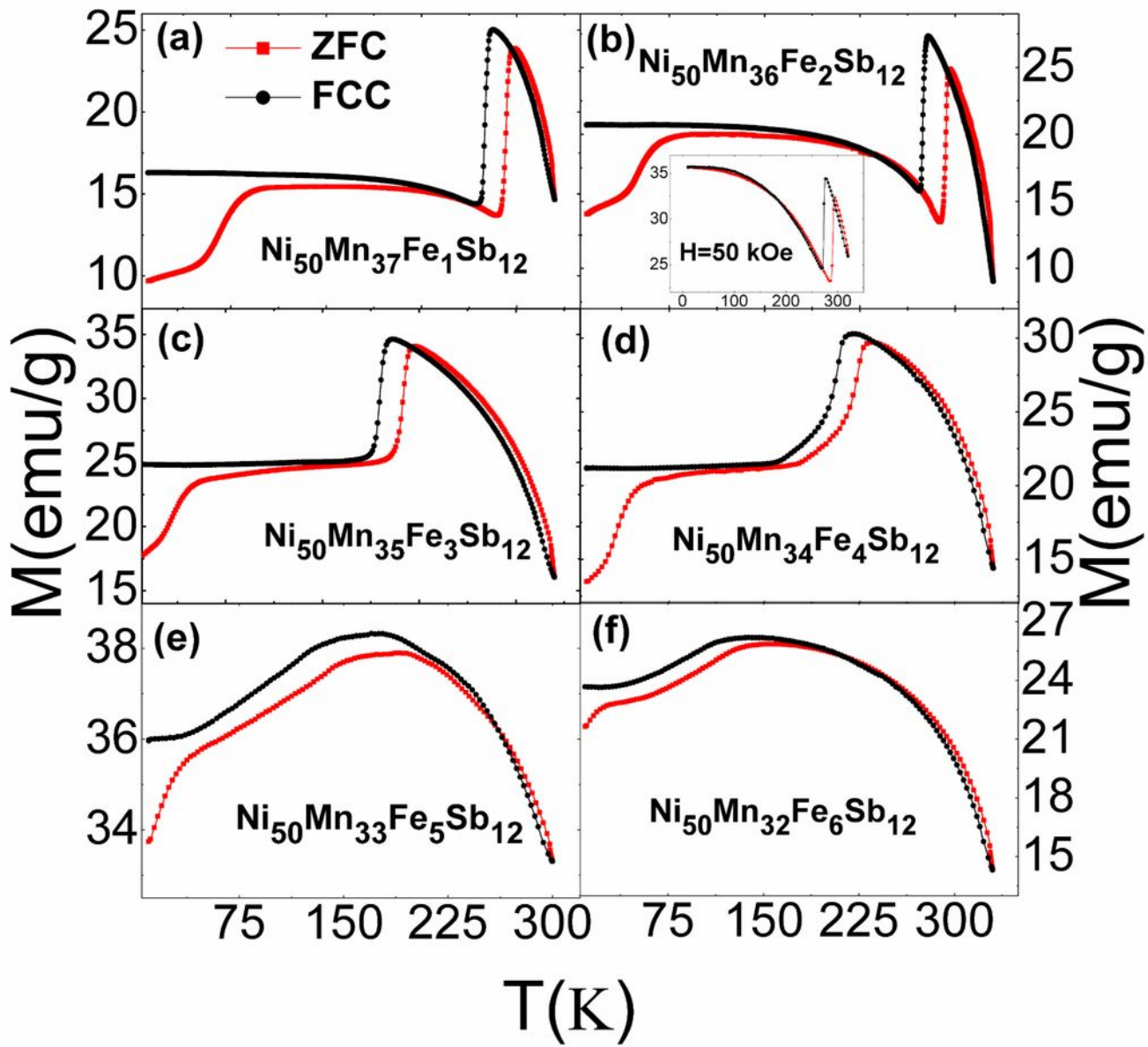

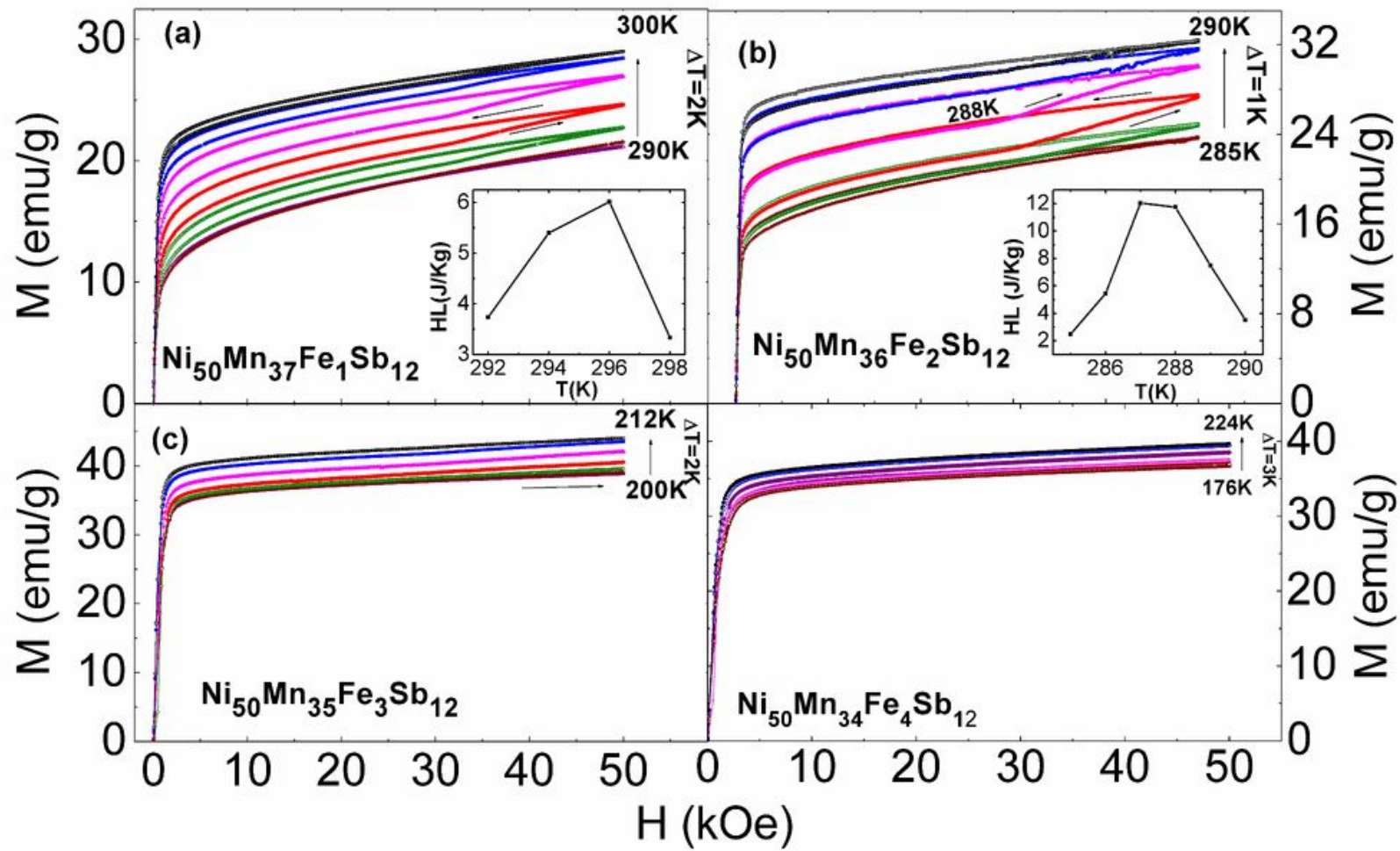

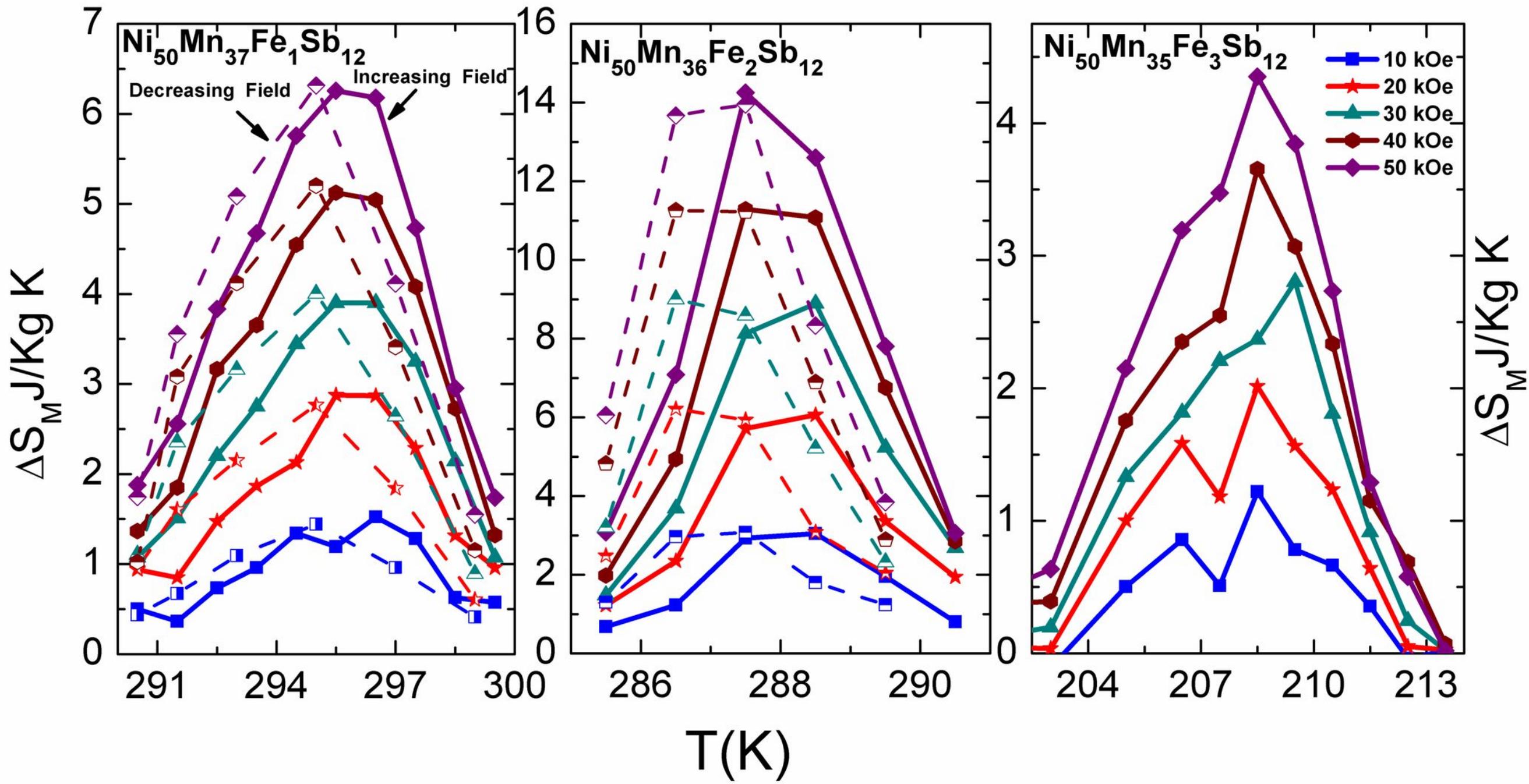

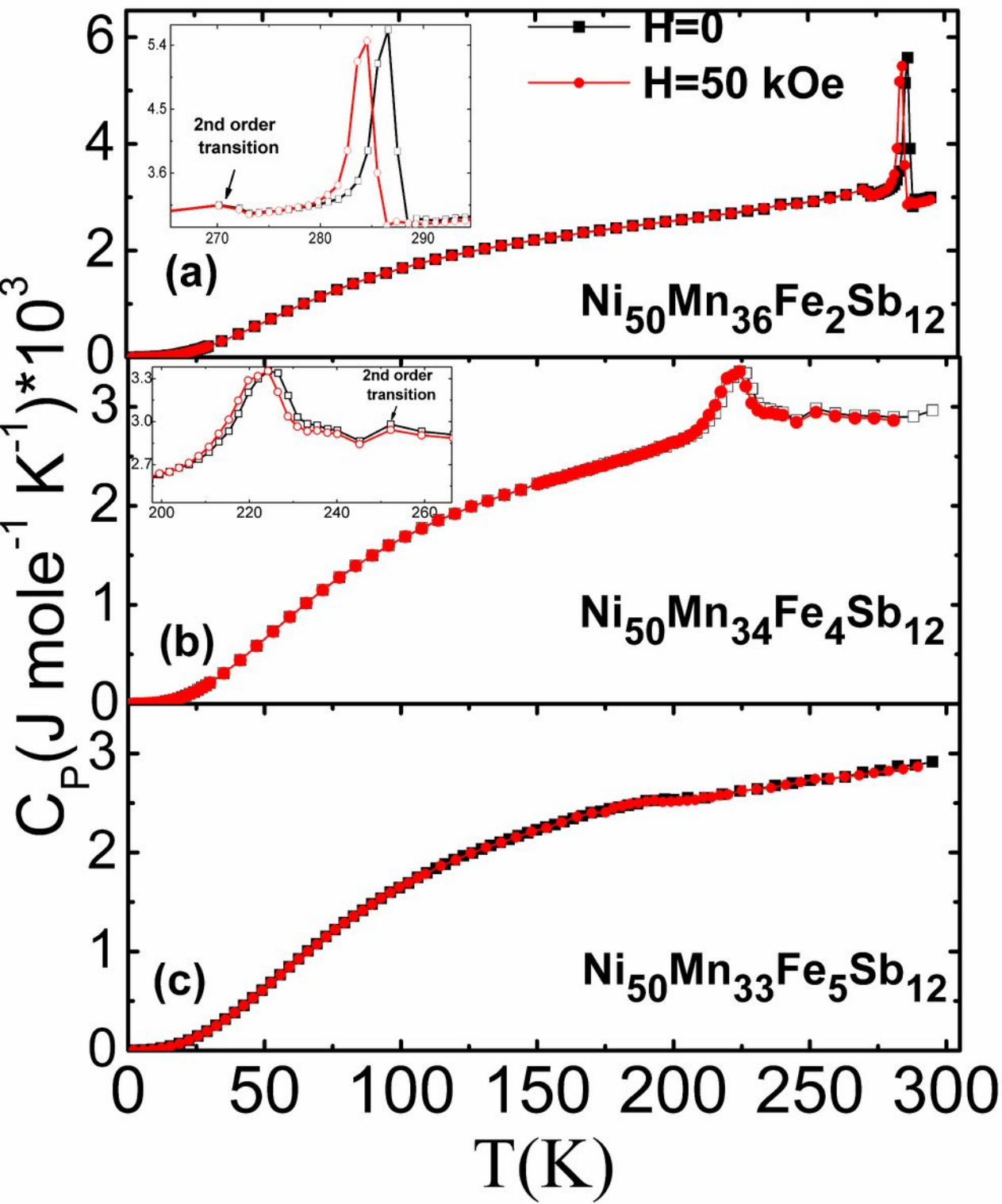

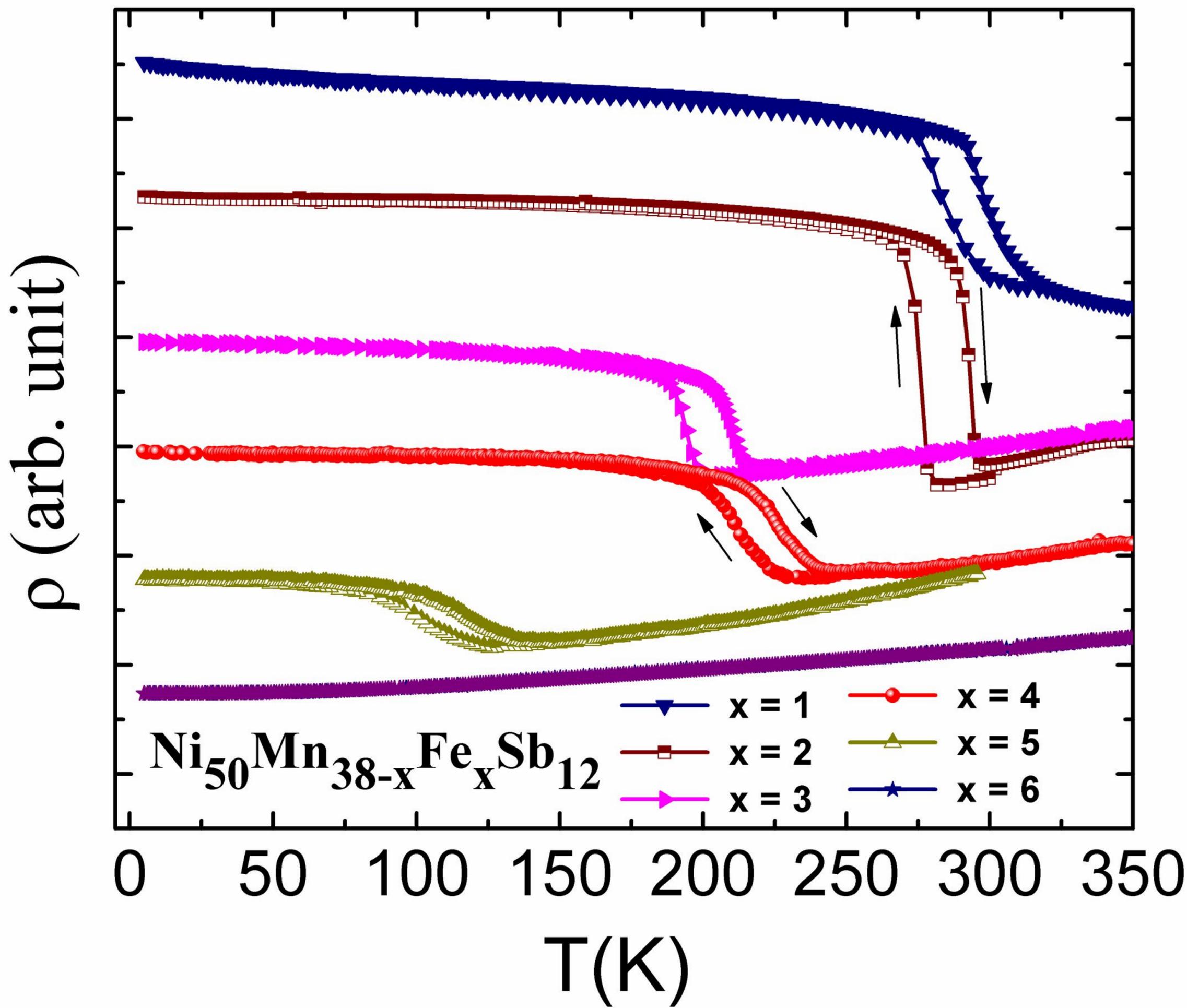

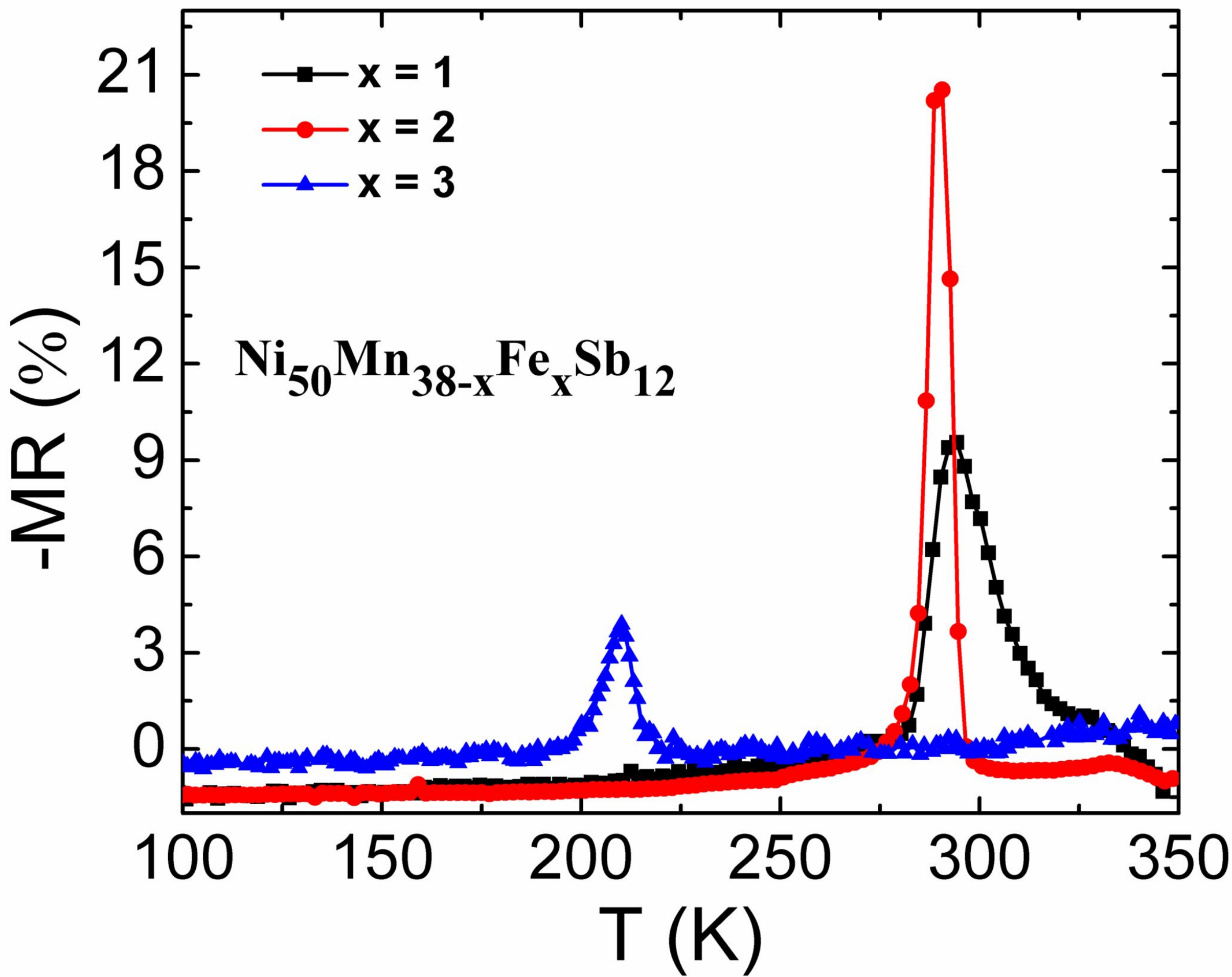

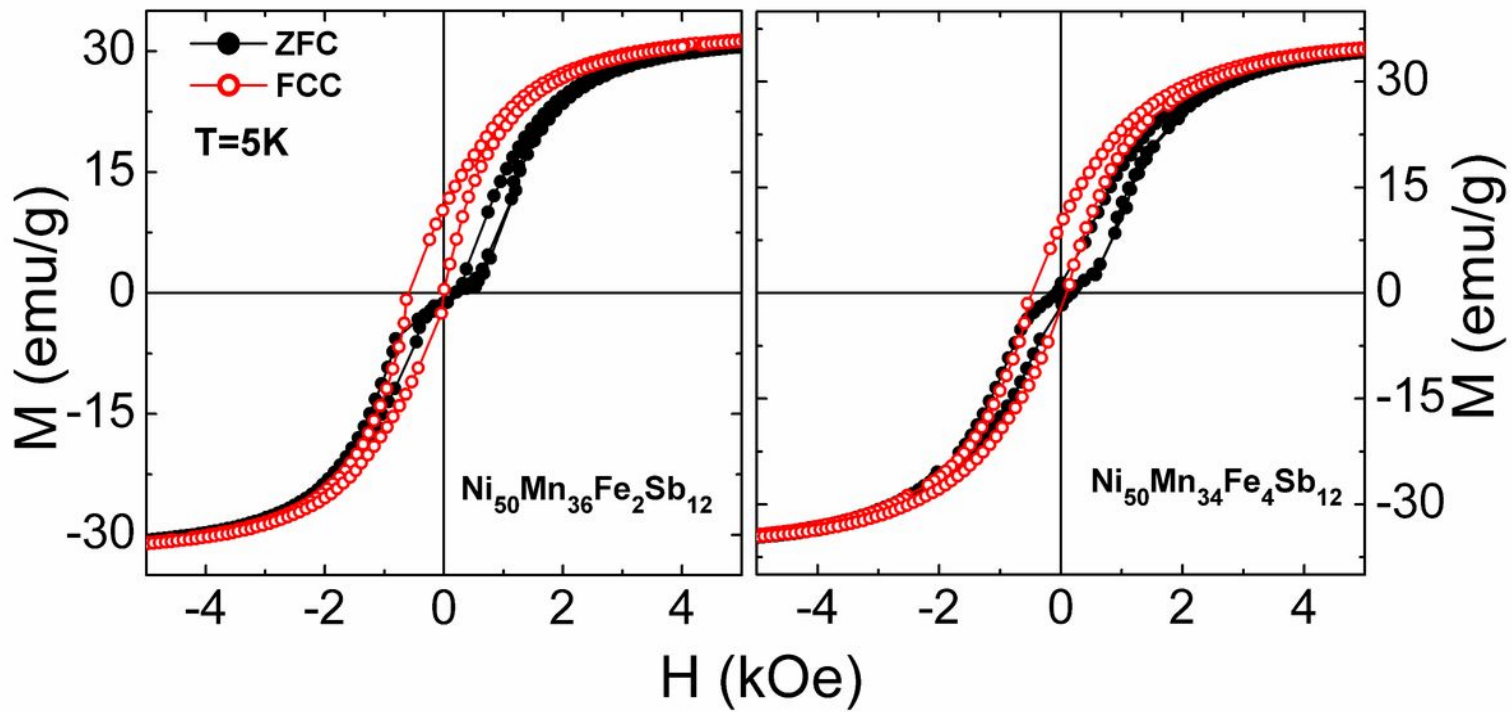

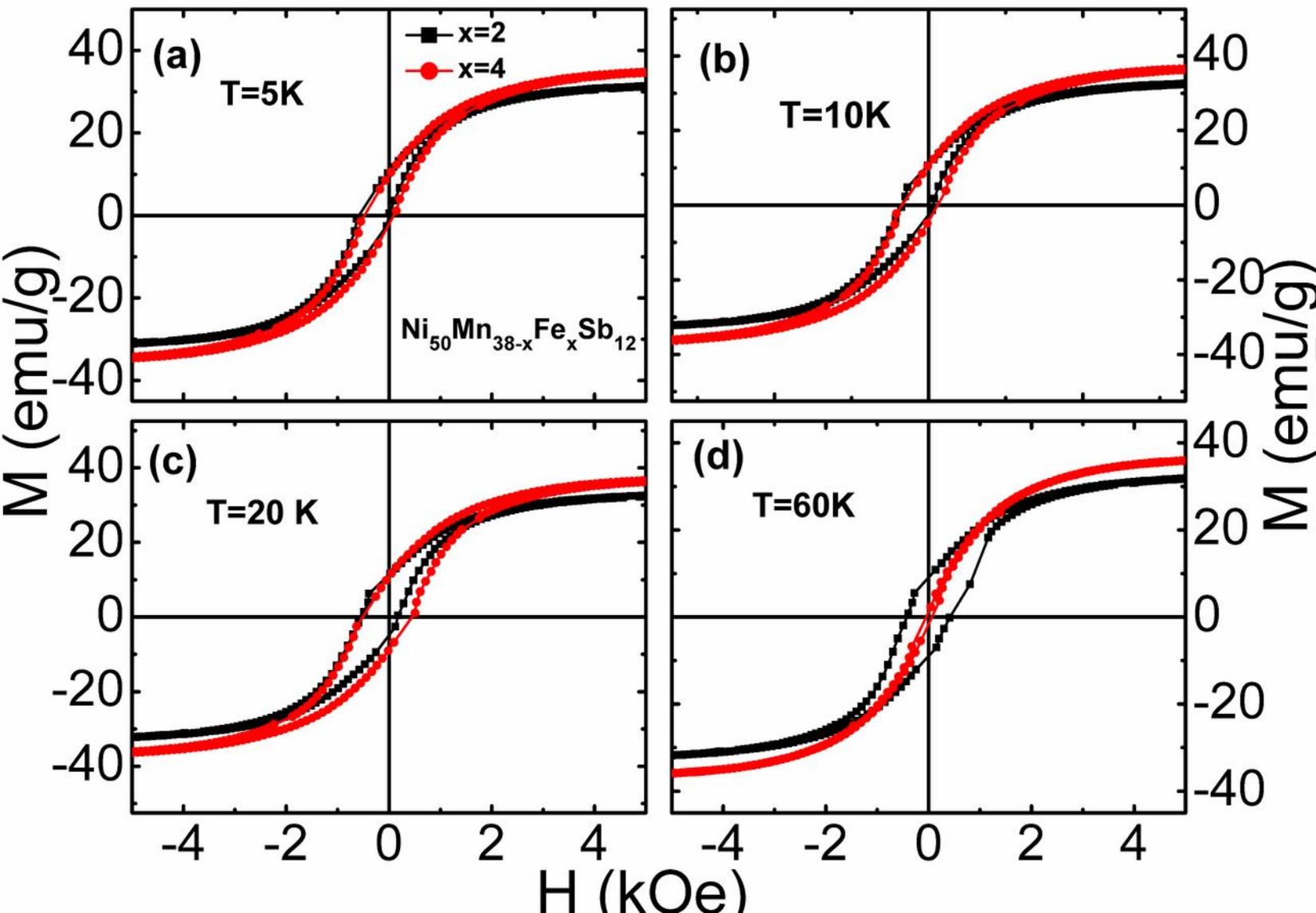

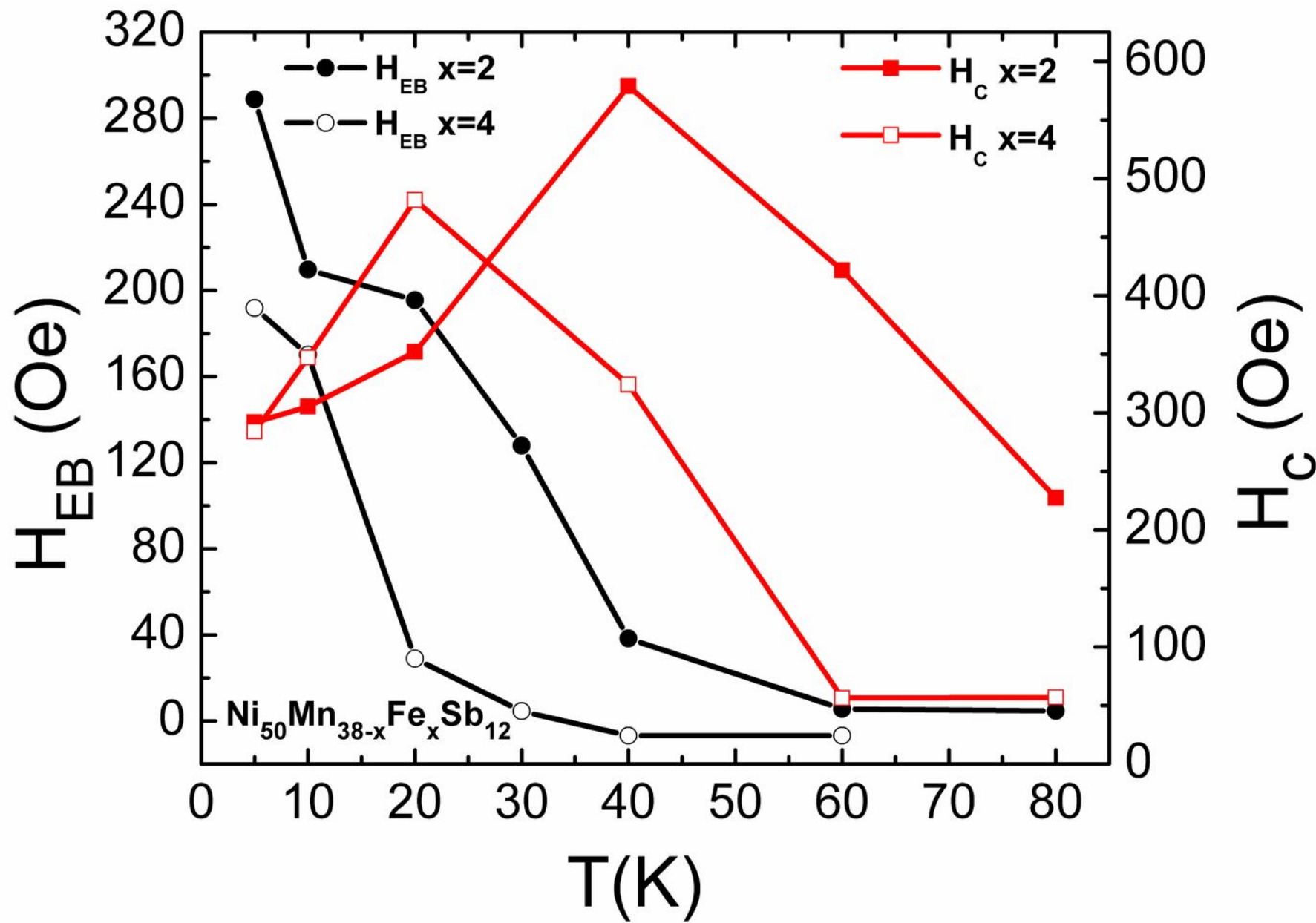